\documentclass[sigconf, authorversion, nonacm]{acmart}

\usepackage{csvsimple}
\usepackage{booktabs} 
\usepackage{siunitx}

\newcommand\vldbavailabilityurl{}
\newcommand\vldbpagestyle{empty} 

\setcopyright{iw3c2w3}

\usepackage{tikz}

\definecolor{myorange}{RGB}{230, 159, 0}
\definecolor{myskyblue}{RGB}{86, 180, 233}
\definecolor{mybluegreen}{RGB}{0, 158, 115}
\definecolor{myvermillion}{RGB}{213, 94, 0}

\usepackage{pgfplots}
\pgfplotsset{compat=1.18}
\usepgfplotslibrary{units} 
\usepackage{cleveref}
\usepackage{subcaption}

\pgfplotsset{
    every axis plot/.append style={
        mark options={solid}
    }
}

\usepackage{listings}
\begin{document}
\title{No Silver Bullet: Boosting GaussDB Performance on the 30TB TPC-H Workload}

\settopmatter{authorsperrow=4}

\author{Tim Zeyl}
\affiliation{%
  \institution{Huawei, Canada}
}

\author{Jason Lam}
\affiliation{%
  \institution{Huawei, Canada}
}

\author{Shu Lin}
\affiliation{%
  \institution{Huawei, Canada}
}

\author{Reza Pournaghi}
\affiliation{%
  \institution{Huawei, Canada}
}

\author{Qi Cheng}
\affiliation{%
  \institution{Huawei, Canada}
}

\author{Calvin Wong}
\affiliation{%
  \institution{Huawei, Canada}
}

\author{Kaixiang Du}
\affiliation{%
  \institution{Huawei, Canada}
}

\author{Yuliang He}
\affiliation{%
  \institution{Huawei, Canada}
}

\author{Yang Sun}
\affiliation{%
  \institution{Huawei, Canada}
}

\author{Weicheng Wang}
\affiliation{%
  \institution{Huawei, Canada}
}

\author{Paul Lee}
\affiliation{%
  \institution{Huawei, Canada}
}

\author{Chen Ruo}
\affiliation{%
  \institution{Huawei, China}
}

\author{Yang Xinyi}
\affiliation{%
  \institution{Huawei, China}
}

\author{Li Qunan}
\affiliation{%
  \institution{Huawei, China}
}

\author{Wang Junjie}
\affiliation{%
  \institution{Huawei, China}
}

\author{Hu Dongxing}
\affiliation{%
  \institution{Huawei, China}
}

\author{Chong Chen}
\affiliation{%
  \institution{Huawei, Canada}
}

\author{Per-\r{A}ke Larson}
\affiliation{%
  \institution{Huawei, Canada}
}

\begin{abstract}

GaussDB is Huawei's premier database system, designed for large-scale deployments and the most demanding workloads. It is a distributed shared-nothing system, capable of handling all types of workloads. This paper outlines a series of modifications to GaussDB aimed at improving its performance on large-scale and complex analytical workloads. After these changes, its performance on the TPC-H workload exceeded the best published result by 40\% at 30 TB. 

The key enhancements to achieve this elite performance include adopting a pipeline execution model, a faster and more scalable inter-node data shuffle, exploiting a unified bus and unified remote memory access. We also expanded the support of cost-based Bloom filter placement and implemented several Bloom filter streaming strategies, enabling their use across nodes.
 \end{abstract}

\maketitle

\pagestyle{\vldbpagestyle}

\ifdefempty{\vldbavailabilityurl}{}{
	\vspace{.3cm}
	\begingroup\small\noindent\raggedright\textbf{PVLDB Artifact Availability:}\\
	The source code, data, and/or other artifacts have been made available at \url{\vldbavailabilityurl}.
	\endgroup
}

\section{Introduction}
\label{sec:introduction}

TPC-H~\cite{tpchWeb} remains one of the most important workloads for assessing analytical queries in relational database management systems. 
It is reported in many papers each year (e.g.,~\cite{lachaud2025understanding,rieger2025t3,tune2025lambda,icde2025whitebox,icde2025weakequiv,otaki2025memory,gro2025linear}) and enterprises rely on it to decide between database systems~\cite{Schneider2025}. 
Vendors and researchers use it to understand specific technological challenges, or \textit{choke points}~\cite{Boncz2014,Dreseler2020a}, that allow them to make targeted improvements that lead to better system performance. In this paper, we summarize the specific improvements we made to GaussDB, Huawei's premier database offering, to achieve elite performance on TPC-H at 30 TB. The TPC-H results presented in this paper remain unaudited at the time of writing, though we believe we have adhered to the detailed benchmark specifications laid out by the Transaction Processing Performance Council (TPC).

Scalability is one of the biggest challenges we faced. As relational databases scale to 
tens of terabytes, database clusters need to scale the number of 
worker nodes that store and process those data. We identified several 
issues that needed to be addressed as we scaled out our clusters
to support larger and larger database sizes.

First, low CPU utilization was often observed during query processing, which we attributed in part to 
our Volcano~\cite{Graefe1994} execution model. This execution model required a large number of threads when running 
under a high degree of parallelism (DOP) in our symmetric multi-processing (SMP) system
on each node. Communication and data exchange across threads require operating system (OS)-level scheduling. 
However, the OS scheduler is not optimized for query processing, relying on mechanisms 
such as pthread condition variables, and leading to a significant CPU overhead for inter-thread communication.
This led us to implement a pipeline execution model (\autoref{sec:pipeline})
that proved to be 
an important factor in maintaining performance at scale.

Second, with na\"ive communication logic for data exchange, we observed poor scalability of network traffic. 
When shuffle is performed, logical communication channels between 
SMP threads on every node must be established with every SMP thread on every 
other node, leading to severe I/O bottlenecks as the number of nodes (or DOP) increases. 
We tackled this challenge
with a more scalable shuffle, described in \autoref{sec:shuffle}, which reduced connections by an exponential factor.

Third, as databases grow, the amount of network traffic
required during shuffle also grows. We found that TCP-based communication 
could lead to network becoming a bottleneck. We realized that we had an opportunity
to speed up network traffic during shuffle by making use of Huawei's 
Unified Bus (UB) communication channels, as described in \autoref{sec:ub}.

Fourth, we further observed that certain strategies for streaming Bloom filters 
across data nodes exhibited poor scaling behavior, so we designed new 
streaming strategies for Bloom filters as described in~\autoref{sec:bloomfilter}. 
We also extended our Bloom filter-aware cost-based optimization~\cite{Zeyl2025} to support distributed deployment. 

Several other challenges were observed, including some sub-optimal plans, which we largely 
addressed by aligning our optimizer's cost model for streaming with 
the above-mentioned improvements to streaming. We also improved statistics 
using HyperLogLog (HLL) synopses to estimate column distinct value counts and
to verify inclusion dependencies in join clauses (\autoref{sec:stats}).

Many of these enhancements are based on known techniques, but have been 
extended and adapted to GaussDB. 
After integrating them, along with several smaller contributions (\autoref{sec:misc}), we achieved 
elite performance on the TPC-H workload.
We measured QphH@30TB = 39,508,107 on TPC-H at 30 TB, 
which is over 40\% higher than the historical published record~\cite{tpchWeb}.
This impressive result was achieved using ARM-based Kunpeng servers, 
while reference systems mainly used Intel-based servers.

The remaining sections of the paper are organized as follows: we present the basic 
GaussDB architecture in~\autoref{sec:architecture}, then we provide details of the key improvements we made to GaussDB. 
We present our pipeline execution model in~\autoref{sec:pipeline}, 
our design for a more scalable shuffle in~\autoref{sec:shuffle}, 
our distributed Bloom filter streaming strategies in~\autoref{sec:bloomfilter}, and our method to detect inclusion dependencies in~\autoref{sec:stats}. 
We present microbenchmark evaluations throughout these sections to highlight the contribution of each, 
and our end-to-end results on TPC-H are presented in~\autoref{sec:evaluation}. All microbenchmark results use a scaled down version 
compared to our full evaluation, but are run using a similar setup.

\section{GaussDB architecture}
\label{sec:architecture}

GaussDB~\cite{Huawei2026,Memarzia2024} is a cloud-native distributed relational database system
based on a shared-nothing architecture. It separates query
coordination, data storage, and execution across multiple nodes while
providing strong transactional consistency and elastic scalability. 
At the cluster level, GaussDB adopts a classic coordinator node (CN) and data node (DN) architecture,
where CNs serve as the SQL entry point, and are responsible
for parsing, optimization, and global execution planning. DNs store data shards and execute query fragments in parallel. A
lightweight centralized transaction management system called the global
transaction manager (GTM) provides globally unique transaction
identifiers and snapshot coordination, ensuring ACID semantics across
shards and nodes. Data are horizontally partitioned across DNs, typically
with replication for fault tolerance, and the system supports intra-availability zone (AZ) or cross-AZ deployment to achieve high
availability and resilience at scale.

\begin{figure}
  \includegraphics[width=\linewidth]{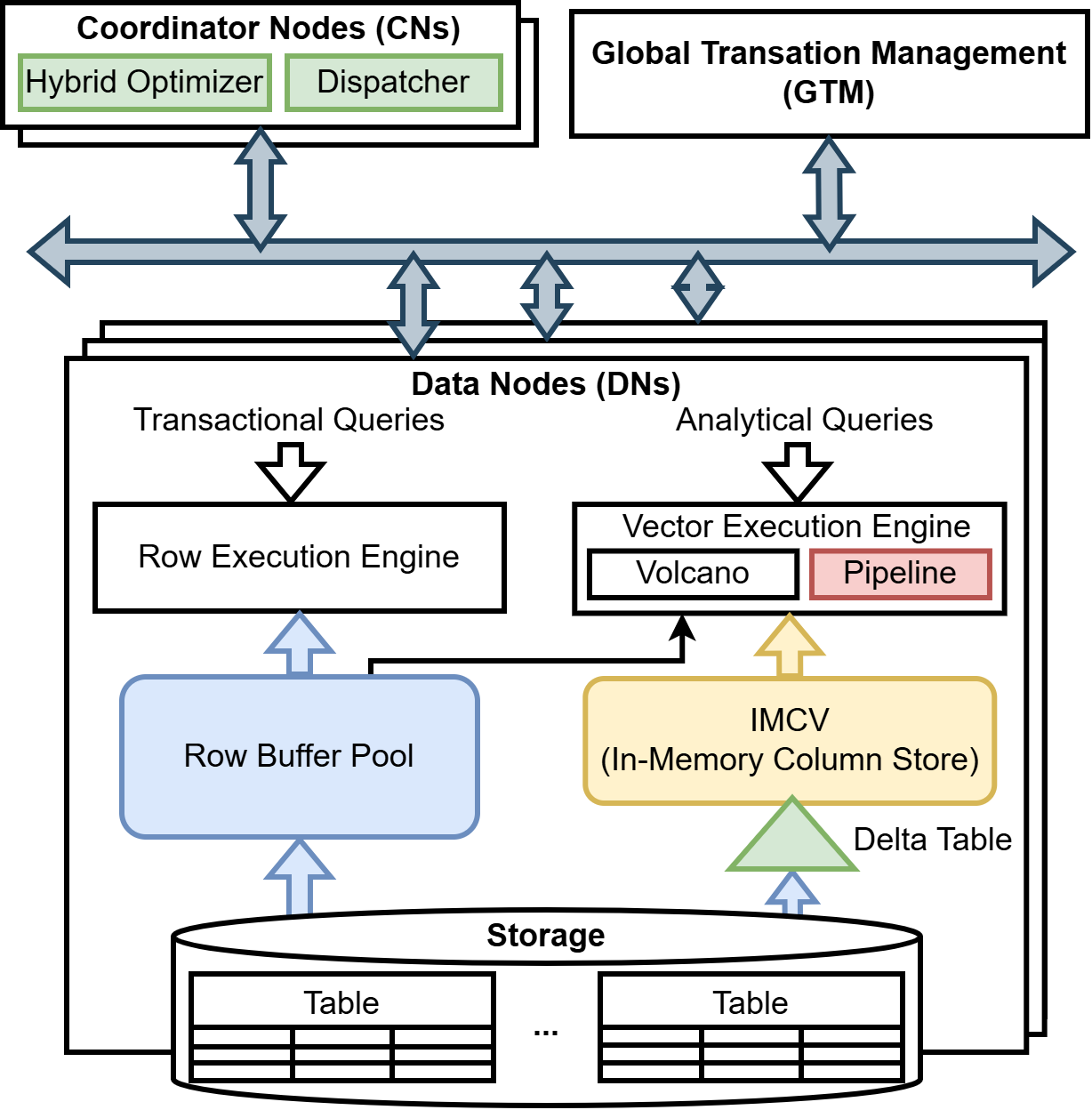}
  \caption{GaussDB distributed HTAP architecture}
  \label{fig:gauss_arch}
\end{figure}

On top of this distributed OLTP foundation, GaussDB's latest release
introduces its HTAP capability through its In-Memory Column View (IMCV)
feature~\cite{Yang2025a,Chen2025}. IMCV is enabled on every DN in the distributed
system. Architecturally, IMCV does not introduce a separate analytical
cluster or external columnar engine; instead, it augments each DN with a
local, in-memory columnar representation that coexists with the primary
row-oriented storage. The row store remains the single source of truth
and persists all data updates, while IMCV serves as a secondary,
non-persistent acceleration structure optimized for analytical scans. 

Within a DN, query execution is split between two specialized engines.
The row execution engine processes transactional queries directly over
row storage and its associated buffer pool, ensuring low latency and
high concurrency for OLTP workloads. In parallel, the vector execution
engine operates over IMCV, executing analytical queries using columnar
layouts and vectorized operators.
Incoming queries are first classified at
the CN and then dispatched to the appropriate execution engine. For
analytical queries, execution plans generated at the CN are decomposed
into subplans and then pushed down to DNs, where the vector engine attempts
to read data directly from IMCV. If the required columns are not present in
IMCV, execution can seamlessly fall back to row storage.

Similar to other widely used HTAP commercial systems (e.g. Oracle~\cite{Lahiri2015},
SQL Server~\cite{Larson2015}), a key component enabling correctness and freshness
is the use of delta tables on each DN. Transactional
updates, such as inserts, updates, and deletes, are applied immediately
to the row store and recorded in delta tables, which logically track
changes relative to the columnar view. During analytical execution,
delta tables are consulted and merged with IMCV data to present a
transactionally consistent snapshot to the vector engine. This mechanism
allows GaussDB to avoid eagerly rewriting columnar data on every update,
while still ensuring that analytical queries observe completely fresh data, 
including even uncommitted updates
within the current transaction. Importantly, IMCV itself is not
persisted to disk; it is a volatile structure that can be reconstructed
from the row store and delta information after restart, simplifying
recovery and preserving the row store as the authoritative data
representation.

Because IMCV is instantiated independently on each DN, Gauss\-DB's
HTAP capability scales in a fully distributed manner. Analytical queries
benefit from both intra-node vectorization and inter-node parallelism,
as columnar scans are executed locally on each DN over its data
partitions, shuffled across multiple DNs, then aggregated through the CN.
This design avoids centralized analytical bottlenecks and aligns
naturally with GaussDB's distributed execution framework. At the same
time, OLTP and OLAP workloads are tightly co-located at the node level,
sharing storage and execution infrastructure, which minimizes data
movement and eliminates the need for asynchronous data replication
between transactional and analytical systems.

By combining a shared-nothing distributed architecture, specialized
execution engines, and per-node in-memory columnar views layered over a
row-native storage engine, GaussDB achieves a scalable and tightly
integrated HTAP system that supports both high-throughput transactions
and low-latency analytics within a single distributed database platform. 

The enhancements described in this paper, in the following sections, are
focused on improving the vector engine, which handles analytical queries. 
As the row engine handles transactional queries, we do not expect our 
enhancements to affect the performance of transactional queries.

\section{Pipeline execution framework}
\label{sec:pipeline}

We implemented a pipeline execution framework, similar to what is described in~\cite{Leis2014}. The new
pipeline SQL engine co-exists with the existing Volcano~\cite{Graefe1994} engine, as shown in \autoref{fig:gauss_arch}. Both engines
retrieve data through common access interfaces that abstract underlying sources, such as row stores, column stores, or
external network input. A key design choice was that the storage layer itself remains agnostic to which execution engine
is used, thereby allowing seamless interoperability, consistent query semantics, and the ability to evolve execution strategies
without modifying the underlying storage components.

We did not at this point change the SQL optimizer for the pipeline engine.  
The optimizer produces a physical plan that is
optimized for the Volcano engine. 
A pipeline builder then decomposes the plan tree into 
linear pipelines as a post-processing step. Each pipeline starts with a 
\textit{source} operator and ends with a \textit{sink} operator; data flow from the source to the sink. 
Pipeline breaker operators 
include materialization operators such as hash-aggregate, sort, and 
hash-join-build, as well as exchange operators, which perform data
redistribution or broadcasting. For example, a \textit{sort} operator in the query 
plan decomposes into a \textit{sort sink} operator in the upstream pipeline and a 
\textit{sort source} operator in the downstream pipeline. Similarly, an exchange
operator breaks into an \textit{exchange sink} operator in the upstream pipeline 
(a.k.a. a producer) and an \textit{exchange source} operator in the downstream 
pipeline (a.k.a. a consumer).

\begin{figure}
  \includegraphics[width=\linewidth]{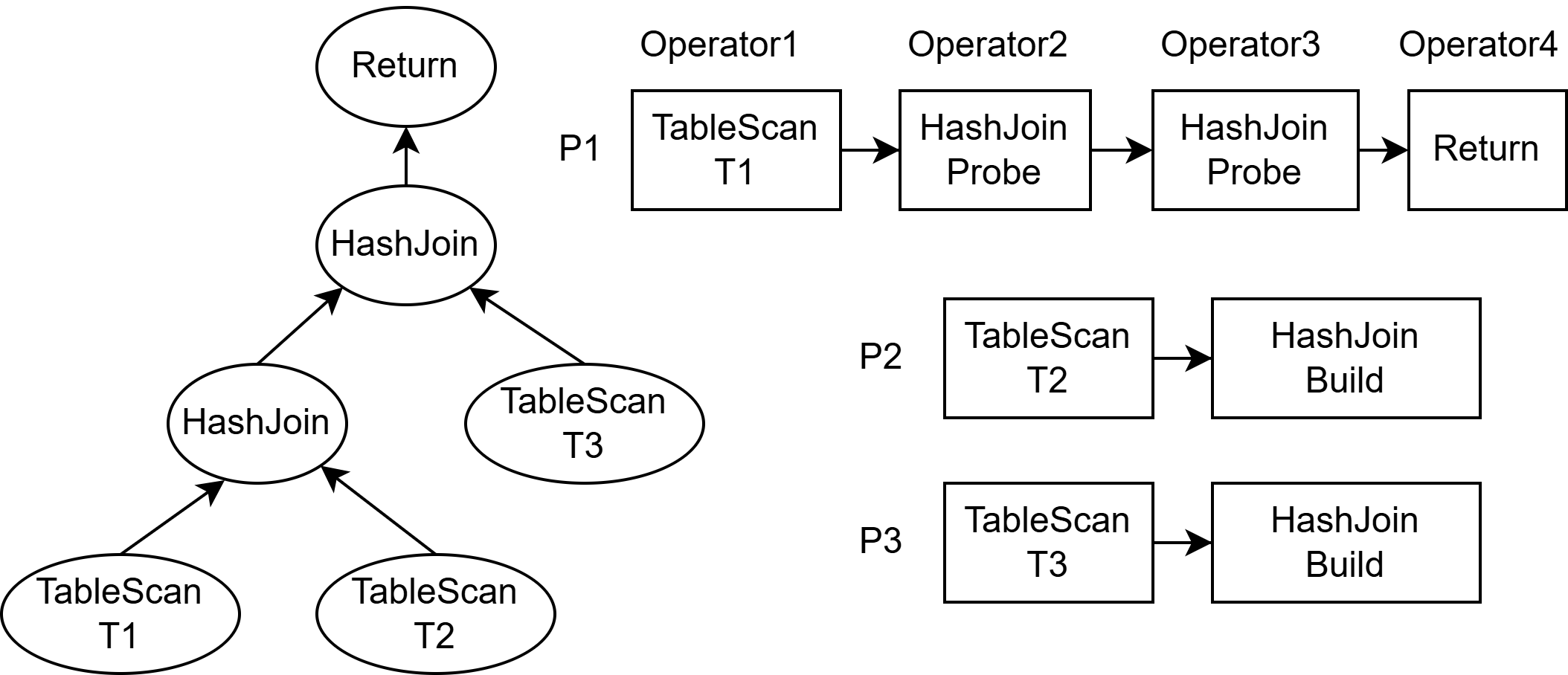}
  \caption{Hash join tree decomposed into pipelines}
  \label{fig:decompose}
\end{figure}

Dividing a plan tree into pipelines opens up opportunities for parallelization that are 
not available under the Volcano model. For example, in \autoref{fig:decompose},
a hash join tree is decomposed into three pipelines, in which pipelines P2 and P3 
can run in parallel if CPU and memory resources are available. 
Each pipeline is divided into $t$ pipeline tasks based on the query DOP. 
Pipeline tasks operate on disjoint data within a pipeline, and data are exchanged at local or remote exchange operators (\autoref{sec:shuffle}). 
The pipeline engine uses $x$ threads to execute all pipeline tasks in each query, where $x$ is determined by the minimum of the number of CPU cores and the 
number of concurrent pipeline tasks in that query. 
For simultaneous queries, this thread model defers some thread scheduling to the OS to avoid the need for custom scheduling logic and to automatically handle thread-level starvation at the cost of some additional context switching overhead.

\subsection{Pull and push model}

To run a pipeline task, we use a driver function, which loops over the operators in the pipeline. 
The driver function calls getOutput() on an operator, and then calls
addInput() on its immediate downstream operator. Within a pipeline, we use a pull model, which means that the driver function first exhausts the output of an operator before pulling data from its immediate upstream operator. For example, for pipeline P1 of \autoref{fig:decompose}, the driver function continually calls operator3's getOutput() function
until the output is exhausted; then the driver pulls from the upstream operator by calling
operator2's getOutput()---if that output is not exhausted, the driver then calls operator3's addInput(). Next, the driver again tries 
operator3's getOutput().
If operator2's output becomes exhausted, the driver calls operator1's getOutput(), followed by operator2's addInput(), etc.
If the source operator has no output, then either the source operator declares it is finished, or the pipeline task is blocked and will be dispatched again when the source operator has output, as described in \autoref{subsection:dispatching}.
The pull model is similar to the Volcano execution engine, in which an operator first exhausts its output, and then it recursively pulls data from child operators. 
In pipeline execution, a loop replaces this less efficient recursion.

Under this pull model, we can efficiently manage memory across the operators of the pipeline. To explain, we denote a batch of data that is passed between operators as a VectorBatch. The memory ownership
of a VectorBatch belongs to the operator that originally allocated it. When addInput() is called on a downstream operator, the \textit{right to use} the memory is passed to that downstream operator. The downstream
operator can modify the VectorBatch memory and pass it further down. When getOutput() is called on an operator, the operator can safely reuse the memory of a VectorBatch that it previously allocated because our pull model guarantees that no downstream operator still needs to access that VectorBatch.

Between pipelines, we use a push model. For pipeline-breaking materialization operators, the downstream 
pipeline is blocked until the upstream pipeline finishes executing. Take the sort operator 
as an example: here the upstream pipeline must finish caching all data in a \textit{sort 
sink} before the downstream pipeline can start pulling data from a \textit{sort source}. Because the entire operation must finish before the downstream can start, it is natural for the upstream pipeline to \textit{push} the downstream pipeline. 

For exchange operators, we use a cache to connect the producer pipeline
and the consumer. The producer pipeline pushes VectorBatches to the cache. As long as the cache has free space, the producer pipeline can continue
to push. The consumer pipeline can execute as long as the cache is not empty. The push model enables the producer pipeline and the consumer pipeline to run in parallel, if CPU resources are available.

\subsection{Local-exchange cache}

Data exchange plays a significant role in query execution. We will discuss remote data exchange in \autoref{sec:shuffle}, but here we describe our implementation of \textit{local} data exchange (i.e., through memory rather than network). Every producer pipeline task has a local exchange cache with up to 64 slots. 
Each slot can store a VectorBatch. To manage slots, we use a 64-bit atomic integer, named \textit{cache\_status}, to indicate the vacant or occupied status of each slot. Each bit in \textit{cache\_status}
corresponds to a slot; a set bit means that the slot is vacant. If \textit{cache\_status} is 0, the cache is fully occupied. Vacant slots are quickly located using a built-in compiler function to count trailing zeros (\_\_builtin\_ctzll), which uses a single highly efficient CPU instruction. 

Each slot has a reference count (also an atomic integer) to track the consumption of VectorBatches. If a VectorBatch
should be sent to N consumers, the reference count is initialized to N. As each of the N consumer pipeline tasks finishes copying data from the producer, they decrement the reference count for that VectorBatch. The consumer that decrements
the reference count to 0 is responsible for setting the \textit{cache\_status} slot to vacant. This single-writer-multi-reader design enables lock-free management of local VectorBatch broadcast: only atomic variables are used to synchronize the readers and writer.

For redistribution, the producer groups the rows in the VectorBatch by their target consumers. For example, rows [1, $n_1$]
should go to consumer1; rows [$n_1$+1, $n_2$] should go to consumer2, etc. This grouping provides memory locality for consumers; when consumers read the data, they have fewer cache misses.

Memory for the cache is dynamically allocated by the producer task as it produces VectorBatches. Once the producer task is finished, the last consumer can release the cache memory. The consumer pipeline task simply tests whether all slot bits in \textit{cache\_status} are 1 to find out whether it is the last consumer. We allocate
cache memory as needed and release it when the exchange is finished to reduce peak memory consumption.

\subsection{Event-driven pipeline dispatching}
\label{subsection:dispatching}

A pipeline execution framework must be able to allow tasks to transition from blocked to ready states, i.e., to perform pipeline dispatching. 
Instead of having a dedicated dispatcher thread, we let pipeline tasks self-dispatch. 
 When a task produces an event that permits
other tasks to transition from a blocked to a ready state, the event-producing task is responsible for performing the state transition of those affected tasks.

There are two kinds of dependencies to consider: an upstream-downstream dependency and a sideways dependency. The upstream-downstream dependency is tracked in both the source operator and the sink operator.
For example, if the local exchange cache is full, then the exchange-sink operator's \textit{acceptInput} status is false. Similarly, if the local exchange cache does not have data for a consumer pipeline task, then
the exchange-source operator's \textit{hasOutput} status is false. The sideways dependency is tracked as a dependency counter at the pipeline level. For example, in \autoref{fig:decompose}, if
Bloom filters are built on both hash joins and pushed to the Scan-T1 operator, then pipeline P1's dependency counter is initialized to four: two originate from the two HashJoinProbe
operators; two originate from the Bloom filters that the Scan-T1 operator needs to apply.

When a pipeline task produces an event that removes a dependency for the target task, the event-producing task checks whether the target task has other dependencies; if not, it can
dispatch the target task (i.e. transition it to ready). The conditions for dispatching a pipeline task are simple: the source operator has output; the sink operator is accepting input; and the sideways dependency counter is 0. Lock-free
techniques are again used to handle race conditions, such as when: 1) multiple pipeline tasks attempt to dispatch the same target pipeline task at the same time; or 2) a running pipeline task \textit{A} is transitioning to a blocked state while another pipeline task is concurrently trying to dispatch \textit{A}.

\begin{figure}
  \begin{center}
    \begin{tikzpicture}
      \begin{axis}[
          name=cpuplot,
          width=\linewidth, 
          height=4.5cm,
          xlabel={Time}, 
          ylabel={CPU Utilization},
          xmin=0,
          xmax=22,
          ymin=0,
          ymax=100,
          y unit=\si{\percent},
          legend pos = north east,
          cycle list={ 
            {myorange, mark=*, thick},
            {myskyblue, mark=square*, thick},
            {myorange, mark=triangle*, dashed, thick},
            {myskyblue, mark=diamond*, dashed, thick}
        }
        ]
        \addplot
        table[x=Time,y=Total,col sep=comma] {data/q21_pipeline_cpu.csv};
        \addlegendentry{PL total}
        \addplot
        table[x=Time,y=Total,col sep=comma] {data/q21_volcano_cpu.csv};
        \addlegendentry{VOL total}
        \addplot
        table[x=Time,y=Sys,col sep=comma] {data/q21_pipeline_cpu.csv};
        \addlegendentry{PL sys}
        \addplot
        table[x=Time,y=Sys,col sep=comma] {data/q21_volcano_cpu.csv};
        \addlegendentry{VOL sys}
      \end{axis}
      \begin{semilogyaxis}[
        at={(cpuplot.south)},
        anchor=north,
        yshift=-1cm,
        width=\linewidth,
        height=3.5cm,
        legend pos = north east,
        xmin=0,
        xmax=22,
        xlabel={Time}, 
        ylabel={Context Switches},
          cycle list={ 
            {myorange, mark=*, thick},
            {myskyblue, mark=square*, thick},
            {myorange, mark=triangle*, dashed, thick},
            {myskyblue, mark=diamond*, dashed, thick}
        }
         ]
         \addplot
         table[x=Time,y=pswitch,col sep=comma] {data/q21_pipeline_pswitch.csv};
         \addlegendentry{PL}
         \addplot
         table[x=Time,y=pswitch,col sep=comma] {data/q21_volcano_pswitch.csv};
         \addlegendentry{VOL}
      \end{semilogyaxis}
    \end{tikzpicture}
    \caption{Running Q21: Pipeline versus Volcano}
    \label{fig:runQ21}
  \end{center}
\end{figure}
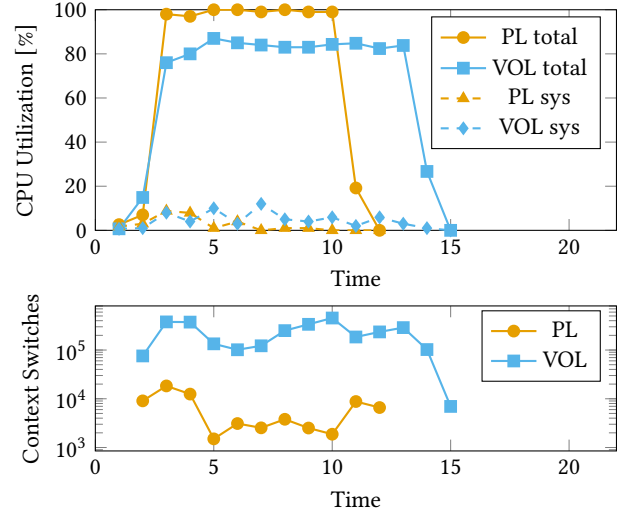

\subsection{Microbenchmark} We demonstrate the advantage of our pipeline engine over our Volcano engine using a single-node 100 GB deployment of TPC-H with DOP=32, running query 21 (Q21). The pipeline engine is \qty{28}{\percent} faster than the Volcano engine when executing Q21.
\autoref{fig:runQ21} shows that during the execution of Q21, the pipeline engine is able to push the CPU to full utilization and uses lower system CPU than the Volcano engine. The pipeline engine uses fewer threads and has a smaller thread synchronization overhead. This is illustrated by the orders-of-magnitude fewer context switches compared to the Volcano engine.

\subsection{Related Work}
The concept of pipeline breakers was introduced by~\cite{Neumann2011}, and they compiled their pipelines into machine code using the LLVM compiler framework. \cite{Leis2014} provided further gains using NUMA-aware task scheduling. We deviated from the push model in these works with our push-pull model that enables efficient memory management. Modern database engines that also employ a pipeline execution model include DuckDB~\cite{duckDb}, StarRocks~\cite{starrocks}, and PrestoDB~\cite{Sethi2019, Sun2023, presto}.
DuckDB supports only a single database node and does not use local exchange for parallelizing queries. Instead, it implements parallelism-aware operators. StarRocks and PrestoDB do support multiple database nodes, and also support local exchange, but our local exchange cache management using built-in compiler functions and atomic variables is unique.

\section{Scalable shuffle}
\label{sec:shuffle}

Shuffle refers to redistributing a table among data nodes, using certain attributes of the table as a redistribution key. Shuffle is a frequent operator in distributed database systems. For
example, in a partitioned distributed join, both join sides must be partitioned on join keys, and shuffle may be required to meet the partitioning requirements.  In an SMP and MPP
deployment, the number of partitions in a shuffle is the number of data nodes multiplied by the DOP on each node. This section describes how we designed a communication system to handle shuffles with high numbers of partitions caused by high DOP and many DNs. 

\subsection{Mailboxes and quotas}

\begin{figure}
  \includegraphics[width=\linewidth]{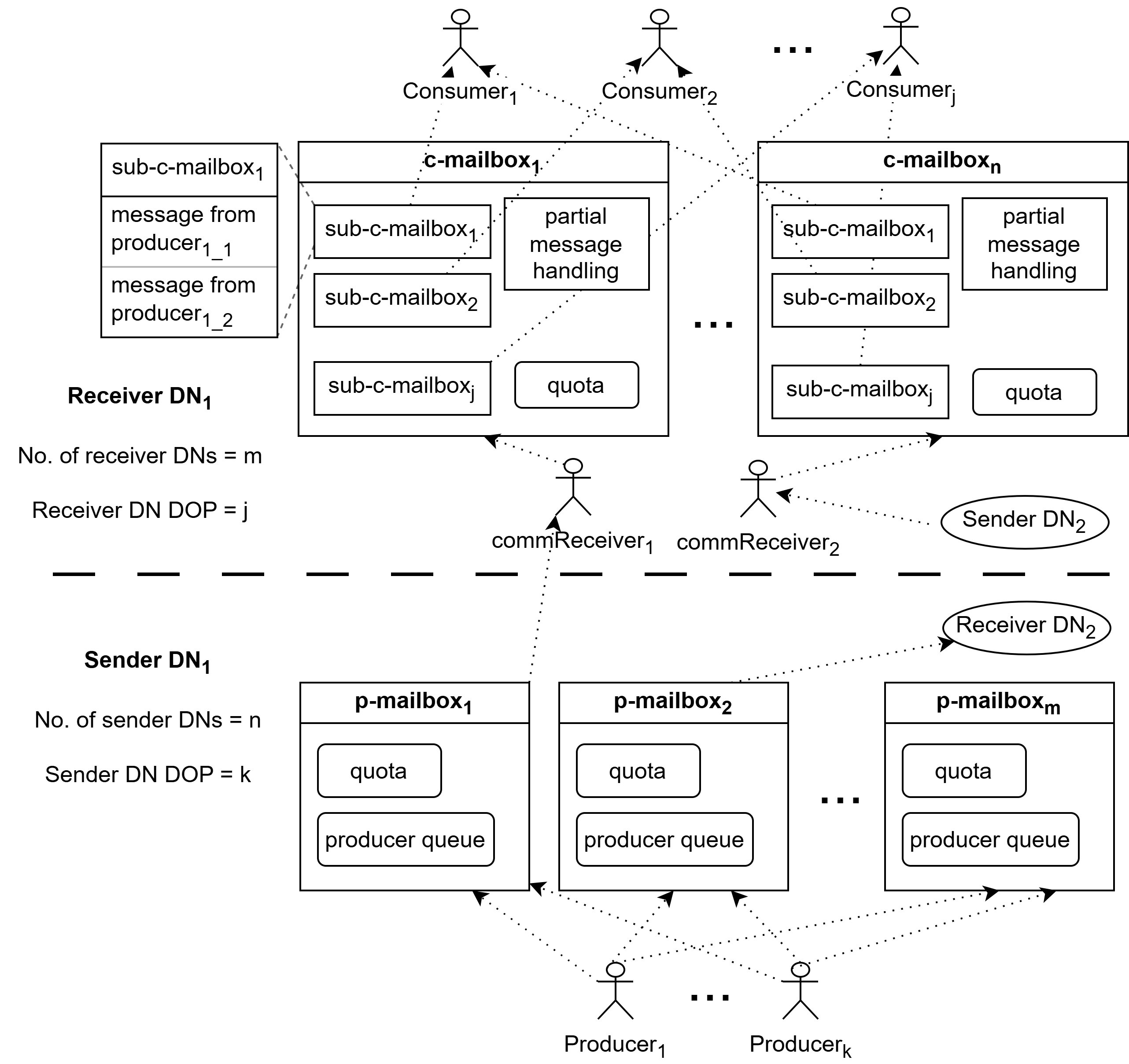}
  \caption{Mailbox system}
  \label{fig:mailbox}
\end{figure}

A core component of the communication system is a set of mailboxes, as illustrated in \autoref{fig:mailbox}. The receiver DN (top) has $n$ c-mailboxes, one for each of the $n$ sender DNs.
The sender DN (bottom) has $m$ p-mailboxes, one for each of the $m$ receiver DNs. A c-mailbox, p-mailbox pair denotes a logical connection between a receiver
DN and a sender DN. There are $\textrm{DOP}=j$ sub-c-mailboxes in a c-mailbox and each sub-c-mailbox stores messages for a consumer partition; a message is a serialized VectorBatch. A consumer task needs to scan $n$ sub-c-mailboxes (one in each c-mailbox) to collect messages. 
Dedicated \textit{commReceivers}, running as background threads, are responsible for receiving messages off-the-wire and delivering them to the sub-c-mailboxes, based on the message header.
The p-mailbox on the sender DN stores quotas, which are mirrored in the corresponding
c-mailbox on the receiver DN. There are $\textrm{DOP=}k$ producer tasks that share the $m$ p-mailboxes. When a producer task wants to send messages to a consumer partition, it must
decrement the quota in the corresponding p-mailbox. If quota has been exhausted, the producer task enters the producer queue in the p-mailbox to wait for quota. Quotas are managed at the DN level.
That is, when there is quota available in a p-mailbox, any producer task can use it to send to the corresponding c-mailbox, regardless of the target sub-c-mailbox. After a consumer task consumes a message, it
evaluates whether the quota should be replenished based on memory availability and the number of unprocessed messages. Note that when a consumer task sends out quotas to a sender DN, it actually sends
out quotas on behalf of all its peer consumer tasks, because quotas are shared at the DN level. This \textit{shared quotas} mechanism reduces the number of quota control messages, and improves scalability.

\subsubsection{Data message buffering}

There are two kinds of messages exchanged between receiver DNs and sender DNs: control messages (e.g., quota control messages) and data messages. Control messages are sent immediately. Data messages
are buffered. We use fixed-size buffers (e.g. 8 KB). A producer task allocates $m \times j$ buffers, corresponding to the total number of consumer partitions. VectorBatches are serialized
into the buffers, and then buffers are sent across the network when they are full.

\subsubsection{Microbenchmark}

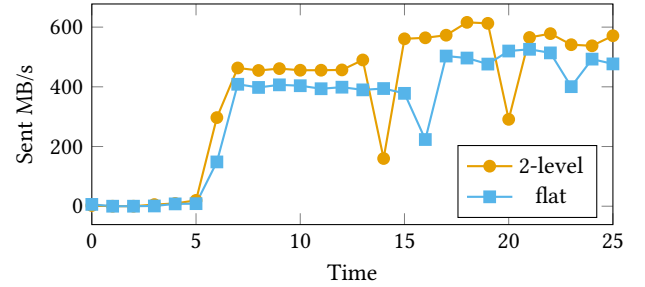
\begin{figure}
  \begin{center}
    \begin{tikzpicture}
      \begin{axis}[
          width=\linewidth, 
          height=4.5cm,
          xlabel={Time}, 
          ylabel={Sent MB/s},
          xmin=0,
          xmax=25,
          legend pos = south east,
          cycle list={ 
            {myorange, mark=*, thick},
            {myskyblue, mark=square*, thick},
            {myorange, mark=triangle*, dashed, thick},
            {myskyblue, mark=diamond*, dashed, thick}
          }
        ]
        \addplot
        table[x=Time,y=write,col sep=comma] {data/q9_750_dop32_commOn.csv};
        \addlegendentry{2-level}
        \addplot
        table[x=Time,y=write,col sep=comma] {data/q9_750_dop32_commOff.csv};
        \addlegendentry{flat}
      \end{axis}
    \end{tikzpicture}
    \caption{Data transfer rate: two-level-mailbox versus flat}
    \label{fig:runQ9}
  \end{center}
\end{figure}

In our initial design, each pair of producer task and consumer task has their own p-mailbox-c-mailbox logical connection. That is, the consumer DN has $n \times k \times j$ c-mailboxes (there are no sub-c-mailboxes),
and the producer DN has $m \times j \times k$ p-mailboxes. We call this design a \textit{flat-mailbox}. 
However, we found that since quotas were managed at the DOP partition level, the flat-mailbox design did not scale well when DOP or the number of DNs gets large. Our two-level-mailbox design manages quotas at the DN level and reduces the number of communication channels needed. The performance difference is demonstrated by testing TPC-H query 9 with various number of DNs and DOP configurations at a scale factor of size 3.75 TB and 7.5 TB. \autoref{tab:mailbox} shows the percentage improvement of query latency using our two-level-mailbox design compared to the flat-mailbox design. The improvement increases when DOP increases and the number of DNs increases. The two-level-mailbox design can achieve a faster data transfer rate than the flat-mailbox design, as shown in \autoref{fig:runQ9} for the configuration DNs=32 and DOP=32.
The transient dips in data transfer rate correspond to the completion of remote exchange operations as the query continues to the next phase of processing. The two-level-mailbox run reaches these transition points earlier than the flat-mailbox run.
\begin{table}[h!]
  \begin{center}
    \caption{Elapsed time comparison: two-level-mailbox versus flat-mailbox}
    \label{tab:mailbox}
    \begin{tabular}{l|S|S|S|S}
      \toprule 
      \textbf{Q9} & \textbf{DOP=4} & \textbf{DOP=8} & \textbf{DOP=16} & \textbf{DOP=32}\\
      \midrule
      DNs=16 (3.75 TB)	& \qty{1.41}{\percent} & \qty{0.46}{\percent}  &	\qty{6.41}{\percent}  &	\qty{18.65}{\percent}\\
      DNs=32 (7.5 TB)	& \qty{1.11}{\percent} &	\qty{1.59}{\percent}  &	\qty{11.84}{\percent}  &	\qty{26.89}{\percent}\\
      \bottomrule
    \end{tabular}
  \end{center}
\end{table}

\subsection{Huawei Unified BUS and URMA}
\label{sec:ub}

In addition to our two-level mailbox design, we saw an opportunity to improve
the network communication between DNs, and potentially expedite shuffle even further.
Huawei has recently proposed a Unified Bus (UB)~\cite{UBBaseSpec} network architecture 
to eliminate inefficiencies in communication between heterogeneous resources (CPUs, GPUs, NPUs, DPUs, etc.) that demand fine-grained coordination---inefficiencies that have resulted from siloed interconnect stacks 
built from layered PCIe, Ethernet, and software-based RPC abstractions. 
Instead, UB elevates memory semantics and 
peer-to-peer communication to the data center scale, enabling direct, 
low-latency, and high-bandwidth interaction among heterogeneous 
resources. We saw an opportunity to make use of this network architecture during remote shuffle in GaussDB.

\subsubsection{UB-based data shuffle optimization in GaussDB}

By default, GaussDB uses TCP-based communication for data shuffle. Between each pair of DNs, two dedicated TCP channels are established, which separately handle read and write traffic. We observed that data shuffle among DNs for some TPC-H queries incurs substantial communication volume and latency which becomes a potential performance bottleneck. We replaced the original TCP channels with UB communication channels to leverage its high concurrency, large bandwidth, and low latency characteristics to improve overall query performance.

Unified Remote Memory Access 
(URMA) is the core low-level communication and data-movement abstraction for UB
that provides RDMA-like remote memory semantics (one-sided, two-sided, 
and atomic operations) across different entities such as CPUs, 
GPUs, NPUs, and DPUs~\cite{UBSoftwareReference}.  
This design eliminates an application-level
 connection setup, supports asynchronous non-blocking execution, and 
allows heterogeneous devices to directly access remote memory without CPU mediation. 
URMA supports several bidirectional communication modes~\cite{UBSoftwareReference} including reliable sequence (a connection-oriented, in-order transmission mode) and reliable messaging (a connectionless but reliable, potentially out-of-order transmission mode). 
Even though logical shuffle streams require in-order transmission, the entire stream between DNs is not required to be in order, so we selected the reliable messaging communication primitive for its better performance, writing our own order-preserving buffer per logical shuffle stream on the receiver side.

Upon GaussDB startup, a configurable number of UB jetties (i.e. long-lived communication ports) 
are established on each DN to exploit parallelism (we use 64). 
URMA memory is registered in advance and shared across all jetties per DN, effectively reducing the overall memory consumption imposed by URMA. 
During shuffle, in our two-level mailbox design, data from multiple producer tasks are collected in each p-mailbox, and the logical stream from each p-mailbox is bound to a dedicated jetty. 
If there are simultaneous shuffle operations, each shuffle operation will have its own set of p-mailboxes, each of which become mapped 
to a jetty using a hashing function.
To transmit, the sender writes metadata, including the source node ID, the stream ID, the data length, the sequence number, and a checksum into the memory header, followed by the payload data. 
The data is transmitted using the URMA send interface, which bypasses the OS kernel and avoids user-kernel context switches. 
Once the data have been successfully copied to the receiver's URMA memory, the jetty generates a completion event and
the associated memory block is returned to the URMA memory pool, enabling efficient memory reuse.

On the receiver side, a dedicated thread continuously polls for completion events, and then the metadata is parsed to extract the node ID, stream ID, and sequence number. If the sequence number is in order, the data are directly enqueued into the corresponding c-mailbox, from which consumer threads asynchronously retrieve and process the data. However, since the reliable messaging mode does not guarantee in-order delivery, we introduce a lightweight order preservation mechanism at the receiver DN.
If an out-of-order message is detected, the data are  temporarily buffered in a reordering queue until all preceding messages arrive, after which the contiguous data sequence is delivered to the c-mailbox.
Empirically, more than 90\% of the messages are received in order, so the reordering overhead is minimal.

\begin{table}[htbp]
\centering
\caption{Speedup by using UB instead of TCP}
\label{tab:tpch-opt}
\begin{tabular}{l r @{\hspace{1.5cm}} l r}
\toprule
\cmidrule(r){1-2} \cmidrule(l){3-4}
TPC-H & \textbf{Speedup} & TPC-H & \textbf{Speedup} \\
\midrule
Q01 & 0.67\%  & Q12 & -2.18\% \\
Q02 & -7.25\%  & Q13 & 16.52\% \\
Q03 & 17.81\%  & Q14 & 4.45\% \\
Q04 & -2.38\% & Q15 & 22.60\% \\
Q05 & 11.41\% & Q16 & 15.68\% \\
Q06 & 0.17\% & Q17 & -4.64\% \\
Q07 & 6.94\%  & Q18 & -5.33\% \\
Q08 & 5.55\%  & Q19 & -8.49\% \\
Q09 & 18.47\% & Q20 & -8.19\% \\
Q10 & -0.12\%  & Q21 & -3.04\% \\
Q11 & -0.42\%  & Q22 & 13.88\% \\
\midrule
\multicolumn{4}{c}{\textbf{Total: 7.18\%}} \\
\bottomrule
\end{tabular}
\end{table}

\subsubsection{Microbenchmark}

We evaluated the effectiveness of UB-based communication using an 8-node GaussDB cluster with TPC-H scale factor 7500  (i.e. 7.5 TB), comparing TCP and UB performance. As shown in \autoref{tab:tpch-opt}, queries with heavy shuffle (Q03, Q05, Q09, Q13, Q15, Q16, and Q22) benefit significantly from UB. Overall, we saw a performance improvement of 7.18\% compared to TCP. However, we also observe performance degradation for a subset of queries (e.g., Q02, Q19, Q20). This regression is primarily attributed to kernel level CPU overhead by enabling URMA communication, especially during memory allocation and reclamation. 
This outweighs the benefits for queries with limited shuffle volume. 
As future work, we plan to introduce an adaptive mechanism that dynamically selects the communication mode based on expected shuffle data volume.

\subsection{Related work}

The exchange operator was introduced for the Volcano model in~\cite{Graefe1990}, 
and is used to enable parallel join in several systems~\cite{Graefe1993, Bellamkonda2013, ibm2023db2}. \citet{Roediger2015} highlight
a scalability problem with this classic exchange operator: that the logical 
connections and associated communication buffers scale with both the number 
of DNs and the DOP. They proposed a hybrid approach that 
employed morsel-driven parallelism locally so network connections could 
scale only with the number of DNs. Our approach maintains the classical 
exchange operation, but still achieves network connection scaling as a function 
of DNs, rather than both DNs and DOP, through our two-level mailbox design.

Several works use RDMA to speed up shuffle~\cite{Roediger2015,Liu2019}, 
or propose alternative architectures to account for new networking
technology~\cite{Binning2015,Zamanian2020}. GaussDB adopts Huawei's UB network to 
achieve fast networking when applied to shuffle on a traditional 
shared-nothing architecture.

\section{Distributed Bloom filters}
\label{sec:bloomfilter}

Bloom filters are widely used in database management systems to facilitate early data reduction by providing an efficient, probabilistic mechanism for row filtering~\cite{Bratbergsengen1984,Mackert1986,Chen1993,Chen1997,Das2015}. 
This minimizes the volume of data subject to downstream processing, thereby improving query performance.

We previously found that including awareness of Bloom filter candidates during cost-based optimization (CBO) enabled improved query plans compared to post-processing application of Bloom filters~\cite{Zeyl2025}. That initial implementation of Bloom filters for GaussDB was restricted to single-node SMP deployments. We have extended the support for Bloom filters to MPP-SMP deployments. This involved adding the cost of streaming Bloom filters across DNs during CBO. The specific cost depends on the estimated Bloom filter size and the partitioning strategy used for BFs, described next.

\subsection{BF streaming strategies}

In our initial single-node SMP deployment we identified and implemented several Bloom filter variants tailored to different streaming strategies. When both the build-side and the probe-side are multi-threaded (DOP > 1) and each thread operates on a partition, we have two cases. First, \textit{partition-aligned} Bloom filters are used when the partitioning of the probe-side relation is aligned with the build-side (i.e., they have the same partitioning key). Here, $n$ partial Bloom filters are created on the build side and each probe-side thread selects the specific partial filter corresponding to its data partition. Second, in \textit{partition-unaligned} cases the probe-side partitioning does not align with the build-side partitioning. In this case, $n$ partial Bloom filters are also created, but the probe-side must either do distributed lookup to select the correct partial filter (i.e., a \textit{distributed Bloom filter}), or must merge all partial Bloom filters into one full \textit{merge Bloom filter}. A \textit{distributed Bloom filter} can only be used if the build-side partitioning column is \emph{available} on the probe-side relation---meaning the planner identifies a known equivalence between a probe-side column and the build-side partitioning column.

In our current MPP-SMP deployment, we have additional parallelism across DNs. In GaussDB, when performing a distributed hash join, the partitioning key across SMP threads is the same as the distribution key across DNs. This alignment allows us to merge all the per-thread partial Bloom filters for each node, giving us node-level partial Bloom filters. These node-level Bloom filters must be serialized and transmitted to all DNs in the cluster. 
On the receiving node, each node-level partial filter can be merged into a single \emph{full} Bloom filter (i.e. an \textit{MPP-merge Bloom filter}), or the partitioning column on the probe-side relation can be used for distributed lookup (i.e. an \textit{MPP-distributed Bloom filter}) of the appropriate node-level partial filter, provided the partitioning column is available.

Notably, distributed Bloom filters (both local and MPP) can be converted into merge Bloom filters, if their bit-vector sizes are identical, using a simple bit-wise OR operator. However, to keep the same target false-positive rate as a full Bloom filter, the size of the partial Bloom filters must be increased proportional to the number of partial Bloom filters being merged. Suppose the size of a node-local Bloom filter, $m$, is determined by

\begin{equation}
\label{eq:bf_size}
    m = -\frac{2n}{\ln{\left(1-\sqrt{\epsilon}\right)}}
\end{equation}

\noindent where $n$ is the estimated number of node-local distinct values, $\epsilon = 0.05$ is our target false positive rate, and we fix the number of hash functions used to two. Now, suppose we had $d$ DNs, each with the same $n$, then to obtain the same false positive rate after merging $d$ partial filters, the bit-vector size should be adjusted to $m' = m \cdot d$.

Consequently, to employ a merge Bloom filter, each partial filter must be $d$ times larger than its counterpart in a distributed Bloom filter, which is decided at build time.
This increased memory footprint of the merge Bloom filter does not present a significant bottleneck in single-node environments. Therefore, in single-node deployments, we prioritize the merge Bloom filter, because probing a single filter is faster than distributed lookup for every probe-side row.

In contrast, for MPP deployments involving $d$ DNs, each node-level filter must be broadcast to all $d$ participants. Consequently, the total network traffic scales at $md^2/8$ bytes for MPP-merge Bloom filters, compared to only $md/8$ bytes for MPP-distributed Bloom filters. This quadratic growth relative to the number of nodes highlights the \emph{scalability limitation} of MPP-merge Bloom filters and leads us to prioritize the \emph{distributed Bloom filter} strategy for MPP deployments---the inverse of our single-node strategy. 
However, merging remains necessary when the partitioning column is missing from the probe side.

To optimize our Bloom filter performance, we implemented a \emph{bucket} Bloom filter that avoids cache-line misses. It uses two hash functions: the first yields a bucket index (where each bucket is the size of a cache-line), and the second yields a position within that bucket. We save on network by transmitting the serialized first hash values across nodes as opposed to full bit-vectors, then modulate the first hash value on the receiver side to determine the second hash value. We also apply LZ4 compression to the transmitted data.

Explicit software pre-fetching was used during probing, so that upcoming bit-vector indices could be fetched from main memory while the probing of previous indices took place, effectively masking memory stall time.

\subsection{Microbenchmark}

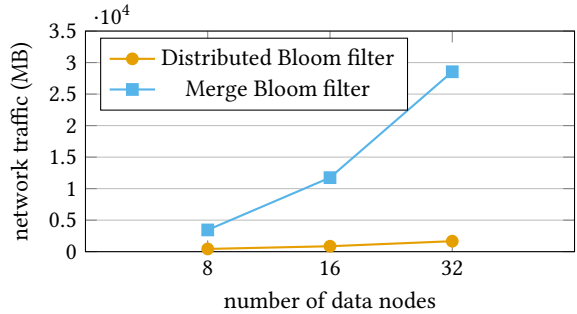
\begin{figure}[ht]
  \centering
    \begin{tikzpicture}
      \begin{axis}[
        width=.95\linewidth, height=4.5cm,
        ymajorgrids=true,
        xmajorgrids=false,
        xmin=0, xmax=4,
         ymin=0, ymax=35000,
        xlabel=number of data nodes,
        ylabel=network traffic (MB),
        xtick={1, 2, 3},
        xticklabels={8, 16, 32},
        ytick distance=5000,
        legend pos=north west,
        cycle list={ 
            {myorange, mark=*, thick},
            {myskyblue, mark=square*, thick},
            {myorange, mark=triangle*, dashed, thick},
            {myskyblue, mark=diamond*, dashed, thick}
          }
      ]
        \addplot coordinates {
          (1, 431.01)
          (2, 862.02)
          (3, 1660.03)
        };
        \addlegendentry{Distributed Bloom filter}
        \addplot coordinates {
          (1, 3448.02)
          (2, 11744.08)
          (3, 28544.28)
        };
        \addlegendentry{Merge Bloom filter}
      \end{axis}
    \end{tikzpicture}
  \caption{Aggregate Bloom filter network traffic and execution time for a subset of TPC-H queries}
  \label{fig:bf_exp_mrg}
\end{figure}

To demonstrate the scalability challenges associated with the MPP-merge filter in a real environment, we conducted evaluations using TPC-H datasets at scale factors (SF) of data size 1.875 TB, 3.75 TB and 7.5 TB.
These were deployed across 8, 16 and 32 DNs, respectively, to maintain a constant data volume per node during cluster expansion.
We selected a subset of TPC-H queries whose plans remained invariant across different scale factors and involved MPP Bloom filters. Figure~\ref{fig:bf_exp_mrg} illustrates the aggregated network traffic as a function of the node count (scale factor), validating the limitations of MPP-merge Bloom filter scalability. Specifically, as the cluster scales, the MPP-merge Bloom filter becomes a performance bottleneck due to its quadratic communication overhead. In contrast, the MPP-distributed Bloom filter maintains superior scalability and outperforms the merge variant at higher node counts.

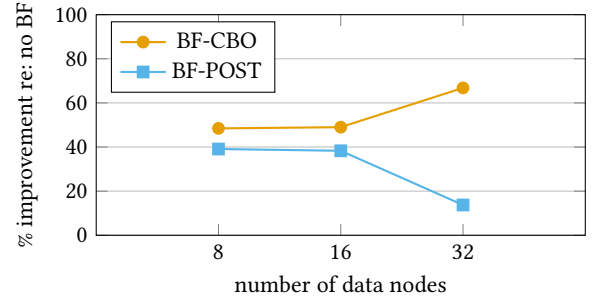
\begin{figure}[ht]
   \centering
   \begin{tikzpicture}
      \begin{axis}[
            width=.95\linewidth, height=4.5cm,
            ymajorgrids=true,
            xmajorgrids=false,
            xmin=0, xmax=4,
            ymin=0, ymax=100,
            xlabel=number of data nodes,
            ylabel=\% improvement re: no BF,
            xtick={1, 2, 3},
            xticklabels={8, 16, 32},
            ytick distance=20,
            legend pos=north west,
            cycle list={ 
            {myorange, mark=*, thick},
            {myskyblue, mark=square*, thick},
            {myorange, mark=triangle*, dashed, thick},
            {myskyblue, mark=diamond*, dashed, thick}
          }
         ]
         \addplot coordinates {
            (1, 48.45)
            (2, 49.00)
            (3, 66.8)
         };
        \addlegendentry{BF-CBO}
        \addplot coordinates {
            (1, 39.1)
            (2, 38.3)
            (3, 13.73)
         };
         \addlegendentry{BF-POST}
      \end{axis}
    \end{tikzpicture}
    \caption{Bloom filter performance in MPP deployments for different TPC-H scale factors}
    \label{fig:bf_overall}
\end{figure}

We also evaluated the broader performance impact of Bloom filters within MPP deployments.
This analysis compares the total execution time of all 22 TPC-H queries, with and without Bloom filters applied, across the same varying data nodes (scale factors).
As shown in Figure~\ref{fig:bf_overall}, placing MPP Bloom filters using a planner post-process (BF-POST) yields an approximate 20-40\% reduction in overall execution time compared to no Bloom filters at all, highlighting the importance of including Bloom filters to reduce data transfer, despite the scalability trade-offs discussed. Including Bloom filter-aware cost-based optimization (BF-CBO) yields an impressive 50-60\% reduction---at higher scales it becomes even more important to ensure the plan is Bloom filter aware.

\subsection{Related Work}

Bit-vector filters have long been used to perform early data reduction in query processing~\cite{Bratbergsengen1984,Mackert1986,Chen1993,Chen1997,Das2015, Bernstein1981} and to limit data transfer over the network in distributed contexts~\cite{Bernstein1981a, Apers1983}. Recently,~\cite{Yang2025b} have discussed various designs for streaming Bloom filters in an MPP environment to achieve predicate transfer~\cite{Yang2023,Zhao2025}. Our BF streaming strategies are most similar to their "Broadcast" based designs, since we only ever stream the build-side data. Yet, our methods differ from their designs in several important ways: first, our environment adds an additional complexity of SMP parallelism at each local node; second, since we always build our BFs on the build-side of a Hash join we can make use of the partitioning column of that join. This allows us to perform distributed lookup or partition-aligned probing of just a single partial filter as opposed to probing all partial filters.

\section{Improving statistics and Foreign key approximation}
\label{sec:stats}

Although database systems allow explicit declaration of foreign key (FK) constraints, in practice, they are often omitted, typically to reduce the overhead on data manipulation operations.
The resulting lack of referential metadata affects the query optimizer's ability to generate valid, alternative execution plans. This challenge has spurred research into techniques to automatically discover FK constraints and inclusion dependencies through data analysis~\cite{Lopes2002,DeMarchi2003,Bauckmann2006,Papenbrock2015,Kruse2017}.

While data analysis cannot infer strict foreign key constraints—as new data may violate them—it can identify approximate relationships, or near-FK constraints. Though lacking the integrity guarantees required for semantic query rewriting, these approximations still provide valuable information for cost-based evaluation and optimization during query rewrite and planning.

We therefore implemented a method for automatic determination of near-FK constraints. This feature enables eager aggregation query rewriting~\cite{Yan1995} (e.g., used in TPC-H query 10 and 13) and prevents the optimizer from adding unnecessary Bloom filters in several queries.

\subsection{Exploiting near-FKs for eager aggregation}
\begin{figure}[ht]
    \centering
    \includegraphics[width=\linewidth]{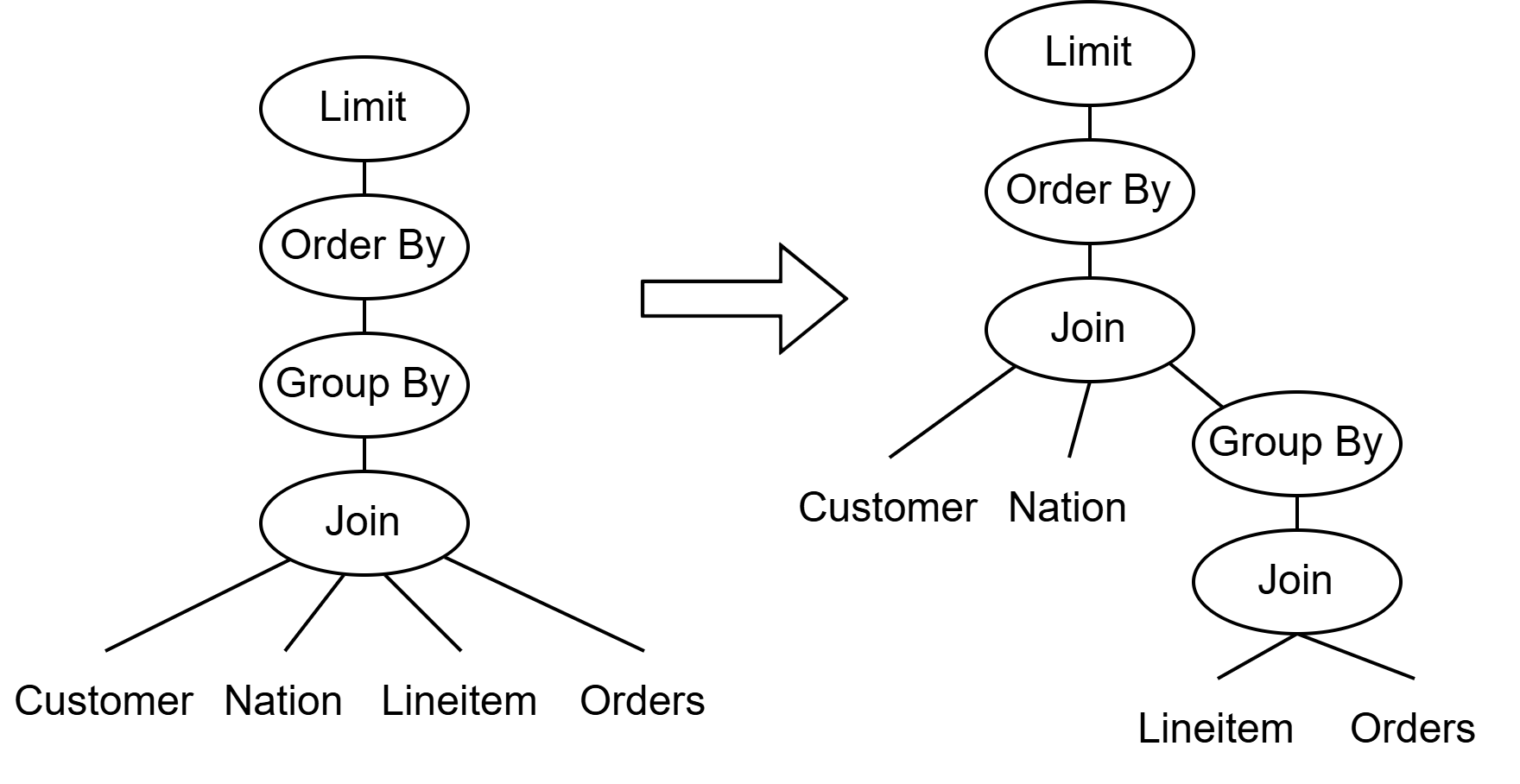}
    \caption{Eager aggregation enabled by near-FKs}
    \label{fig:eagerAgg}
\end{figure}

For TPC-H query 10, under the conditions that \path{c_custkey}, \path{n_nationkey}, and \path{o_orderkey} are primary keys, we can use eager aggregation to push down the GROUP-BY, as shown in \autoref{fig:eagerAgg}, improving performance. However, this optimization is only beneficial when the aggregation cost is not prohibitive. A key determinant of this cost is the selectivity of the downstream join; if it filters out a significant portion of rows, the early aggregation overhead may outweigh the cost of the original query tree. By leveraging near-FK constraints, we can estimate this join selectivity before the CBO phase and determine whether to push the aggregation through the join.

\subsection{Using near-FKs for Bloom filter pruning}
A second application for near-FK constraints is pruning unnecessary Bloom filters in hash joins. When a valid primary-foreign key relationship exists---where the build key is an unfiltered primary key (PK) and the probe key is an FK---the Bloom filter cannot filter any probe rows, as every FK value exists in the PK. Therefore, constructing the filter incurs runtime overhead without benefit. By detecting an approximate PK-FK relationship, we can identify and eliminate these redundant Bloom filters during CBO, thereby reducing the search space and preventing their generation at runtime.

\subsection{Detecting near-FK constraints}
Our method leverages HLL synopses as statistical metadata collected from tables, similar to~\cite{Nazi2018}. But since we are only interested in approximate inclusion dependency detection (instead of the more general inclusion coefficient estimation or FK-PK discovery), we can employ a simplified, lower-cost procedure.

To construct an HLL synopsis, each value in the column is hashed to a value represented as a bit-vector. 
A subset of those bits is used to assign the value to one of $M$ buckets, and the number of trailing zeros, $\rho$, is computed from the remaining bits, with the HLL synopsis keeping track of the maximum number of trailing zeros, $\rho'$, observed in each bucket after scanning all rows. 
It is well known that these $\rho'$ can be used to estimate the number of distinct values (NDV) in each column. 
The intuition is that if $\rho$ trailing zeros are observed in the hash value, it is likely that around $2^\rho$ values have been seen so far, since every bit in the hash value can be thought of as an independent Bernoulli trial.

Now, if column A is an FK of column B, a PK, then the values of A are a subset of the values of B. It follows that, in the $i^\textrm{th}$ bucket in the HLL synopsis of A, the maximum length of trailing zeros $\rho'_{A,i}$ must be less than or equal to the maximum length of trailing zeros in the $i^\textrm{th}$ bucket in the HLL synopsis of B, i.e., 

\begin{equation}
\label{eq:hll_fk_condition}
\forall i \in [1,M], \rho'_{A,i} \le \rho'_{B,i} .
\end{equation}

Our method for determining near-FK relationships is simply to verify \autoref{eq:hll_fk_condition} when encountering a candidate join clause. By construction, this condition avoids false negatives in which we fail to detect a true FK-PK relationship. However, it is susceptible to false positives. False positives occur when we test whether column A is a candidate FK of column B and we find \autoref{eq:hll_fk_condition} holds true even though the values of column A are not in fact a subset of column B.

We can reason about the probability of false positives as follows. In a single HLL bucket with a maximum length of trailing zeros $\rho'_{B,i}$, the probability that a single random hash value has fewer (or equal to) trailing zeros is $p(\rho \le \rho'_{B,i}) = 1 - \frac{1}{2^{\rho'_{B,i}+1}}$ according to the properties of Bernoulli trials. Then, over $M$ buckets and supposing that the candidate column A has $k$ distinct values in each bucket, the probability that we identify a random column as being an FK of B is

\begin{equation*}
\prod^{M}_{i = 1} (1 - \frac{1}{2^{\rho_{B,i} + 1}})^{k}.
\end{equation*}

\noindent This probability increases when the NDV of A is much smaller than the NDV of B, so special handling is required when $k$ is small. However, with a sufficient number of buckets (we use $M=1024$), and by ensuring that column B is a true PK, we find that the probability of false positives remains low. A non-zero false positive rate is tolerable because our use cases 
allow for near-FK constraints rather than strict constraints.

Our detection algorithm itself is efficient. It requires only M comparisons between two HLL synopses (with additional handling of HLL shards) and executes in tens to hundreds of microseconds. Among the join clauses encountered during planning of a TPC-H 7.5 TB deployment, we found zero false positives and zero false negatives.

\subsection{Related Work}

The near-FK relationships we detect in our system are a type of inclusion dependency, a well-studied area in databases~\cite{Lopes2002,DeMarchi2003,Bauckmann2006,Papenbrock2015,Kruse2017}. 
Many works in the area aim to discover all inclusion dependencies from a database schema~\cite{DeMarchi2003, Bauckmann2006}, or an interesting subset thereof~\cite{Lopes2002}, while our method is only used for verification when a join clause is encountered in some planning contexts. Because our method is used during planning, it needs to be fast. Recent work introduces fast~\cite{Kruse2017} and scalable~\cite{Tan2025} methods to find \textit{approximate} inclusion dependencies, which align with our use case, but these approaches either use sampling or partitioning---processes that are still too heavy to run during planning. We found using HLL synopses to approximate inclusion dependencies to be very fast and efficient, though perhaps not precise enough for discovery of inclusion dependencies across a database schema. \citet{Nazi2018} proposed a method that makes use of HLL synopses to estimate the more general inclusion coefficient (a process that involves an iterative binary search), then compared that coefficient to a threshold to identify inclusion dependencies. Our work only requires approximate inclusion dependencies, which allows us to forego the full estimate of the inclusion coefficient and directly compare the HLL synopses of two columns to get an extremely fast approximate estimate.

\section{Additional improvements}
\label{sec:misc}

Beyond the improvements discussed so far, this section explores several additional small improvements that also contributed to our overall performance.

\subsection{SIMD optimization}
Vectorized execution is broadly adopted in database systems. SIMD (Single Instruction, Multiple Data) instructions perform vector operations using hardware registers and instruction sets to execute computations in parallel. Our approach to implement SIMD instructions was to write code that is amenable to auto-vectorization via modern compilers (e.g., gcc, clang), which, if configured, try to generate vectorized instructions when possible. This saves effort and improves maintainability compared to manually producing platform-dependent intrinsics. 
Different hardware platforms have their own vector instruction sets (e.g., the AVX-family for Intel); since ARM-based CPUs were used in our environment, the instruction sets generated in our deployments were SVE/SVE2.

We re-wrote code to take advantage of vectorization in the following areas: 1) expression evaluation, 2) hashing for joins and aggregations, and 3) hashing for Bloom filters. 
In addition, we modified our LLVM code generation, which we use for some input-specific expression evaluations, to produce vectorized execution instructions. 

\subsection{Numeric type optimization}
\textit{Numeric} is a fixed-point number type in GaussDB, where a maximum \textit{precision} and \textit{scale} are defined during type declaration. Precision in this context is the maximum number of digits in the number, while scale is the number of digits after the decimal point. This type is common in database tables, for example, to model currency, interest rate, or when exact precision is required, as in TPC-H.

We updated GaussDB's handling of \emph{Numeric} such that table columns of this type could be stored 
in vector form, by using an integer type to store their values. We require all values in the vector 
to have a common scale, then we can remove the decimal point and store the value as an integer---the 
scale is stored separately.
This representation allows us to perform basic arithmetic using integer instructions, with certain constraints.
Addition and subtraction can be performed between vectors with the same scale; the resulting vector 
will have the same scale as the operands. When the scale does not match, the vector with smaller scale 
can be rescaled before the operation. Multiplication can be performed with vectors 
having different scales; the scale of the resulting vector is based on the sum of the operands' 
scales.  Cases of overflow or underflow need special handling, for example, casting to a wider 
underlying integer type, or reverting to less efficient representations.

\subsection{IMCV compression}

In-Memory Column View (IMCV) supports efficient lightweight compression that is automatically used to keep data size down without suffering high costs for compression and decompression. IMCV compression supports algorithms such as delta encoding, dictionary compression, RLE, LZ4, and ZSTD, and can automatically select compression algorithms based on data characteristics during data loading. 
Compression not only reduces the memory footprint, but counterintuitively it can also improve performance for some queries, especially when using techniques such as late reading and operating on compressed vectors when possible.

IMCV mainly stores column data in memory, but there is also metadata information such as row id, variable-length offset array, and transaction metadata. This makes the memory expansion relatively large when decompressed. At 7.5 TB TPC-H, the IMCV in-memory size across all nodes is measured to be approximately 10 TB; through lightweight compression, the in-memory size is reduced to approximately 4 TB, representing ~60\% reduction. We found that the end-to-end latency on all TPC-H queries had negligible difference compared to without compression at this scale factor.

\section{Evaluation}
\label{sec:evaluation}

\subsection{Setup}
\label{sec:setup}

\emph{Environment}: Evaluation was performed on Huawei's bare metal servers (BMS), using Kunpeng 920 7280Z processors running Huawei Cloud EulerOS (HCE) 2.0 operating system, each with 160 CPU cores (1 vCPU per core), 2048 GB RAM, and 5 TB of SSD storage provided by the elastic volume service (EVS). The servers were interconnected through a UB network.

\emph{Workload}: Our primary evaluation is on the TPC-H workload, an industry-standard decision support workload developed by the TPC.  
TPC-H specifies a read-heavy, complex analytical workload that simulates real-world business intelligence scenarios, using 22 analytical SQL query templates and 2 refresh functions (RFs) on a standardized database schema. Scaling of TPC-H is indicated by a scale factor (SF) with a larger SF value specifying a larger raw data size. An SF=1 is approximately 1 GB of data.

The overall performance metric of TPC-H is called the composite queries-per-hour (QphH@Size), which is derived from a geometric mean of metrics from two tests: 1) a power test and 2) a throughput test.  The power test executes all 22 queries and two RFs in a single stream in a defined order; this measures the fastest speed for single-user performance. The throughput test is performed in multiple parallel streams, where each stream executes 22 queries and 2 RFs in a defined order. SF is a multiplicative factor in the definition of the power test metric and the throughput test metric, so perfectly scalable systems should increase their QphH linearly with SF.

\begin{figure}
  \includegraphics[width=0.7\linewidth]{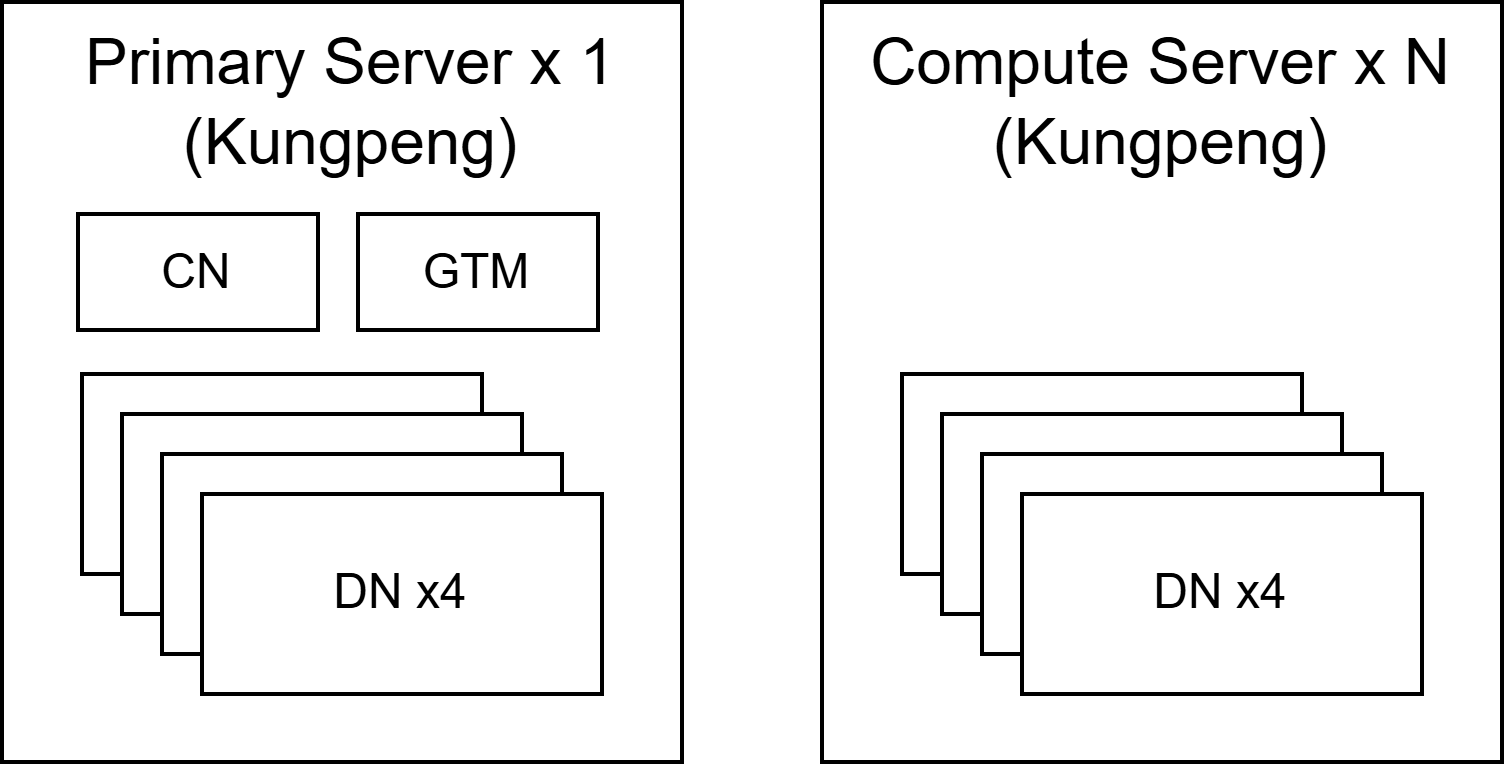}
  \caption{Deployment view of GaussDB DNs, CN and GTM}
  \label{fig:deployment}
\end{figure}

In our deployment, one of the BMS servers was the primary server and the others were compute servers. We deployed a single CN, a single GTM and four DNs on the primary server, and four DNs on each compute server as shown in \autoref{fig:deployment}. The load was expected to be handled primarily by DNs, while the CN and GTM were very lightly loaded, so it was feasible for us to include four DNs on the primary server as well.  In a shared-nothing system, each DN executes its assigned plan fragment independently, so we bound each DN to one of four available NUMA nodes on each server; this isolated CPU and memory usage and avoided potential cross-NUMA memory access overhead.

\subsection{TPC-H Results}

We evaluated GaussDB on TPC-H at SF=30,000 (30 TB) where we deployed 32 BMS servers, resulting in 128 DNs, 64 TB total memory, and 5120 CPU cores (1 vCPU per core).
The overall performance metrics of the power test, throughput test and composite QphH@30TB are shown in \autoref{tab:30t-results} and illustrated for simplicity in \autoref{fig:30t-results}, with the top performing reference systems published on the TPC website~\cite{tpchWeb} also shown for comparison. We include systems with expired results in our comparison, as there was only one reference system with current results at the time of writing.
We remind the reader that our results remain unaudited at the time of writing.

Before starting this work, GaussDB performance was well below the best published TPC-H results. After incorporating all these enhancements, we achieved a composite QphH@30TB of 39,508,107, which is 40\% better than the composite score of Hologres, the next best reference system. 
To our knowledge, this is the first time an ARM-based system is competitive with Intel based systems on TPC-H results. 
Hologres used 11000 vCPUs (physical cores undisclosed), so their composite per-vCPU performance can be computed as 2533 QphH@30TB/vCPU, while our composite performance can be computed as 7716 QphH@30TB/vCPU, which is more than 3 times larger. 
Exasol used 2304 vCPUs (18 servers x 2 CPUs x 32 cores x 2 hyper-threads per core), so they still achieve the highest per-vCPU performance at 9837 QphH@30TB/vCPU.

\begin{table}[h!] 
    \centering
    \begin{filecontents*}{data/30t-overall-transpose.csv}
System,Power,Throughput,Composite
GaussDB,39269930.7,39747729.1,39508107
Hologres,21468670.0,36169338.3,27865885
Exasol,24086987.1,21326631.4,22664825
OceanBase,14788059.7,15757953.6,15265305
\end{filecontents*}
\csvautobooktabular{data/30t-overall-transpose.csv}
    \caption{QphH of reference systems at 30 TB}
    \label{tab:30t-results}
\end{table}

\pgfplotstableread[col sep=comma]{data/30t-overall.csv}{\overallthirty}
\begin{figure}
\centering
\begin{tikzpicture}
    \begin{axis}[
        ylabel={QphH@30TB},
        ybar,                 
        bar width=10pt,       
        symbolic x coords={Power, Throughput, Composite}, 
        xtick=data,           
        legend image code/.code={
          \draw[#1] (0cm,-0.1cm) rectangle (0.3cm,0.1cm);
        },
        legend style={at={(0.5,-0.2)}, anchor=north,legend columns=-1}, 
        ymin=0,      
        enlarge x limits=0.25,
        cycle list={ 
            {fill=myorange,draw=black!80},
            {fill=myskyblue,draw=black!80},
            {fill=mybluegreen,draw=black!80},
            {fill=myvermillion,draw=black!80}
        }
    ]
    \addplot table[x=Test, y=GaussDB]{\overallthirty};
    \addlegendentry{GaussDB}
    \addplot table[x=Test, y=Hologres]{\overallthirty};
    \addlegendentry{Hologres}
    \addplot table[x=Test, y=Exasol]{\overallthirty};
    \addlegendentry{Exasol}
    \addplot table[x=Test, y=OceanBase]{\overallthirty};
    \addlegendentry{OceanBase}
    \end{axis}
\end{tikzpicture}
\caption{QphH of reference systems at 30 TB}
\label{fig:30t-results}
\end{figure}
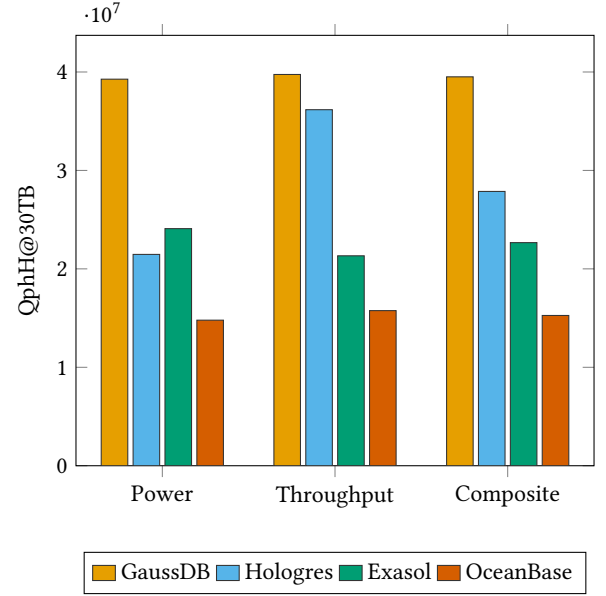

We have a relatively higher percentage improvement over the next best reference system in the power test compared to the throughput test. 
We speculate that this may be due in part to our usage of fewer vCPUs (5120 vs. 11000 for Hologres) and less memory (64 TB vs 88 TB for Hologres), which may be relatively more constrained in our deployment under the high concurrency of the throughput test.

Individual query latencies during the power test are shown in \autoref{tab:30t-query-results}. 
GaussDB has the best performance on most queries and generally performs similarly to the best performing system on the remaining queries. 
GaussDB performs particularly well on several query types, including queries that:
\begin{itemize}
\item make use of the local exchange cache (\autoref{sec:pipeline}), e.g. queries 4 and 12;
\item require remote shuffle (\autoref{sec:shuffle}) or place many Bloom filters (\autoref{sec:bloomfilter}), e.g. queries 2, 7, 8, 9, and 20; and
\item take advantage of near-FK determination (\autoref{sec:stats}), e.g. query 10 and 13.
\end{itemize}

\noindent These observations validate the importance of our improvements to GaussDB.

\begin{table}[h!] 
    \centering 
    \begin{filecontents*}{data/30t-query.csv}
,GaussDB,Hologres,Exasol,OceanBase
Q1,2.89,2.73,8.32,10.22
Q2,1.15,2.65,2.84,1.95
Q3,4.82,4.06,6.52,5.63
Q4,0.94,2.20,1.89,4.43
Q5,3.49,4.88,7.57,13.15
Q6,0.21,0.48,1.40,1.26
Q7,2.73,5.36,10.57,14.36
Q8,3.46,6.26,3.60,12.32
Q9,12.80,31.52,29.27,36.68
Q10,2.26,6.08,6.49,9.98
Q11,20.26,18.44,20.28,9.64
Q12,0.59,2.64,3.22,5.91
Q13,5.73,8.87,8.29,13.36
Q14,1.01,3.34,4.96,5.79
Q15,2.41,1.77,6.16,4.69
Q16,2.69,2.04,6.04,5.58
Q17,1.39,3.69,0.99,7.25
Q18,7.78,11.93,9.01,8.72
Q19,1.40,4.27,1.92,5.63
Q20,4.65,8.69,4.92,11.69
Q21,4.70,12.62,4.22,16.05
Q22,2.07,3.57,0.84,6.89
RF1,5.30,17.32,8.85,6.16
RF2,7.14,10.68,0.74,3.04
\end{filecontents*}
\csvreader[
      tabular=lrrrr, 
      table head=\toprule & GaussDB & Hologres & Exasol &OceanBase \\\midrule,
      late after last line=\\\bottomrule
    ]{data/30t-query.csv}{1=\colone, 2=\coltwo, 3=\colthree, 4=\colfour, 5=\colfive}{%
      \colone & \coltwo & \colthree & \colfour & \colfive
    }
    \caption{Power test query latencies (s) at 30 TB} 
    \label{tab:30t-query-results} 
\end{table}

\section{Conclusion}
\label{sec:conclusion}

GaussDB has evolved over the past few years to become an attractive 
option for enterprises requiring a reliable and performant HTAP 
database. The improvements we have made to our analytical query processing,
including our new pipeline engine, a more scalable shuffle, 
distributed cost-based Bloom filter support, and near-FK detection
among others, 
have allowed GaussDB to achieve elite performance on 
TPC-H at scale.

\balance
\bibliographystyle{ACM-Reference-Format}
\bibliography{tpchpaper}

\end{document}